\documentclass{article}
\usepackage{spconf,amsmath,graphicx}

\usepackage{enumitem}
\setlist{nosep, leftmargin=14pt}

\usepackage{mwe} 
\usepackage{amsthm}
\usepackage{algorithmicx,algorithm}
\usepackage{amsbsy}
\usepackage{amsfonts}
\usepackage{marvosym}
\usepackage{amsmath}
\usepackage{threeparttable}
\usepackage{booktabs}
\usepackage{multirow}
\usepackage{float}
\usepackage{url}
\usepackage{makecell}
\usepackage{bm}

\title{SFCNeXt: a simple fully convolutional network for effective brain age estimation with small sample size}
\name{\begin{tabular}{c}Yu Fu$^{1,2}$, Yanyan Huang$^{1}$, Shunjie Dong$^{1}$, Yalin Wang$^{3}$, Tianbai Yu$^{1}$, Meng Niu$^{4}$ and Cheng Zhuo$^{1,5}$\sthanks{Corresponding author.}\end{tabular}}\footnotesize

\address{$^{1}$Zhejiang University, Hangzhou, China\\
$^{2}$Binjiang Institute of Zhejiang University, Hangzhou, China\\
$^{3}$Arizona State University, Tempe, USA\\
$^{4}$The First Hospital of Lanzhou University, Lanzhou, China\\
$^{5}$Key Laboratory of Collaborative Sensing and Autonomous Unmanned Systems \\ of Zhejiang Province, Hangzhou, China\\}\footnotesize

\begin{document}
%
\maketitle
\begin{abstract}
Deep neural networks (DNN) have been designed to predict the chronological age of a healthy brain from T1-weighted magnetic resonance images (T1 MRIs), and the predicted brain age could serve as a valuable biomarker for the early detection of development-related or aging-related disorders. Recent DNN models for brain age estimations usually rely too much on large sample sizes and complex network structures for multi-stage feature refinement. However, in clinical application scenarios, researchers usually cannot obtain thousands or tens of thousands of MRIs in each data center for thorough training of these complex models. This paper proposes a simple fully convolutional network (SFCNeXt) for brain age estimation in small-sized cohorts with biased age distributions. The SFCNeXt consists of Single Pathway Encoded ConvNeXt (SPEC) and Hybrid Ranking Loss (HRL), aiming to estimate brain ages in a lightweight way with a sufficient exploration of MRI, age, and ranking features of each batch of subjects. Experimental results demonstrate the superiority and efficiency of our approach.
\end{abstract}

\vspace{-0.3cm}

\section{Introduction}
\label{sec:intro}
Brain development and aging, accompanied by complex biological and neuroanatomical changes, is an ongoing and lifelong process that is largely not fully understood~\cite{bashyam2020mri,elliott2021brain,fu2021cross,an2021synergistic}. To capture the development and aging patterns, a brain age estimation model can be trained using healthy neuroimaging data to predict the brain age as closely as feasible to the actual chronological age. The discrepancy between the anticipated age and the actual chronological age is commonly referred to as the “brain age gap”, which has been associated with a variety of biological and cognitive characteristics~\cite{cheng2021brain,he2021multi,fu2022otfpf}. Measuring the difference in brain age of patient groups may help assess the disease heterogeneity and improve disease risk screening~\cite{mori2022brain}. 

T1-weighted magnetic resonance images (T1 MRIs), possess rich morphological information about the brain, have been frequently used for brain age estimations~\cite{cheng2021brain,he2021multi,peng2021accurate}. Whether a model can obtain a smaller mean absolute error (MAE), a larger Pearson correlation coefficient (PCC), and a larger Spearman’s rank correlation coefficient (SRCC) is crucial for determining its suitability for brain age estimation~\cite{he2021multi,fu2022otfpf,peng2021accurate}. Researchers in the deep learning community have built different neural network backbones for brain age estimation with brain T1 MRIs, such as convolutional neural network (CNN)-based~\cite{peng2021accurate,fu2022otfpf}, ResNet-based~\cite{bashyam2020mri,he2021multi} and Transformer-based~\cite{he2022deep}. However, most of these models own complex frameworks (e.g., many branches for information flows) and consume a lot of GPU resources, which may limit their deployment in clinical scenarios. Besides, a good performance of these models is largely depended on huge sample sizes (e.g., thousands or tens of thousands of MRIs), but researchers in a specific team often do not have sufficient MRIs to train, given the considerations of data sharing and privacy issues of multi-site data around the world~\cite{white2022data}.

To address the above issues, we propose a simple, fully convolutional network (SFCNeXt) for accurate and effective brain age estimation with a small sample size. The SFCNeXt consists of two technological parts: Single Pathway Encoded ConvNeXt (SPEC) and Hybrid Ranking Loss (HRL). The SPEC includes a backbone pathway that consists of a four-stage ConvNeXt, and a branch pathway adopts sex features. After the concatenation of the two pathways, a conformer encoder cascaded with a multilayer perceptron (MLP) is employed for adaptive feature fusion for brain age estimation. The HRL is a combination of mean square error (MSE) loss and fast differentiable ranking loss, which not only fits brain MRI features to brain ages but explores the soft ranking relationships of each batch of brain MRIs.

\vspace{-0.3cm}

\section{Datasets and preprocessing}
\label{sec:data}
We evaluate the proposed SFCNeXt on a healthy cohort~\cite{yahata2016small,tanaka2021multi}, which includes T1-weighted MRIs from 11 consortium sites of Japan (see Table~\ref{datasets}), with a total of 779 subjects aged 19-72 years. To our knowledge, this is a small-sized cohort relative to sample sizes adopted in mainstream brain age estimation models (e.g., tens of thousands of samples). The histogram matching and intensity normalization are performed for the harmonization of MRIs across datasets. All images are processed via a standard preprocessing pipeline with FSL 6.0~\cite{jenkinson2012fsl}, which includes nonlinear registration to standard 2mm MNI space and brain extraction~\cite{smith2002fast,fu2022otfpf}. After preprocessing, each MRI has a spatial size of 91$\times$109$\times$91~voxel with a spatial resolution of 2~mm$^3$.

\begin{table}[t]\footnotesize
 \centering
 \caption{Demographic information of brain age estimation datasets used in this study.}
 \label{datasets}
 \setlength{\tabcolsep}{0.9mm}{
 \begin{threeparttable}
 \begin{tabular}{lcccc}  
  \toprule   
  Site & $N_{img}$ & Age Range & \makecell[c]{Age \\ (Mean$\pm$STD)} & \makecell[c]{Gender \\ (Male/Female)}\\
  \midrule   
  ATT & 13 & 20-25 & 22.23$\pm$1.42 & 12/1\\  
  ATV & 39 & 20-30 & 22.69$\pm$2.20 & 29/10 \\  
  CIN & 39 & 20-67 & 38.69$\pm$13.49 & 25/14 \\
  COI & 119 & 20-72 & 50.87$\pm$12.78 & 42/77 \\
  HKH & 29 & 28-65 & 45.41$\pm$9.53 & 12/17 \\
  HRC & 49 & 23-68 & 41.69$\pm$11.66 & 13/36 \\
  HUH & 67 & 20-66 & 34.75$\pm$12.97 & 29/38 \\
  KTT & 75 & 19-57 & 28.89$\pm$9.05 & 48/27 \\
  KUT & 157 & 19-68 & 36.36$\pm$13.18 & 92/65 \\
  SWA & 101 & 19-55 & 28.44$\pm$7.86 & 86/15 \\
  UTO & 91 & 24-72 & 45.31$\pm$14.11 & 31/60 \\
  \midrule
  Overall & 779 & 19-72 & 37.61$\pm$14.32 & 419/360 \\
  \bottomrule  
 \end{tabular}
 \end{threeparttable}}
 \vspace{-15pt}
\end{table}

\vspace{-0.4cm}

\section{Methology}
\label{sec:metho}

\subsection{Single Pathway Encoded ConvNeXt (SPEC)}
\begin{figure*}[tb]
 \centering
 \centerline{\includegraphics[width=13.3cm]{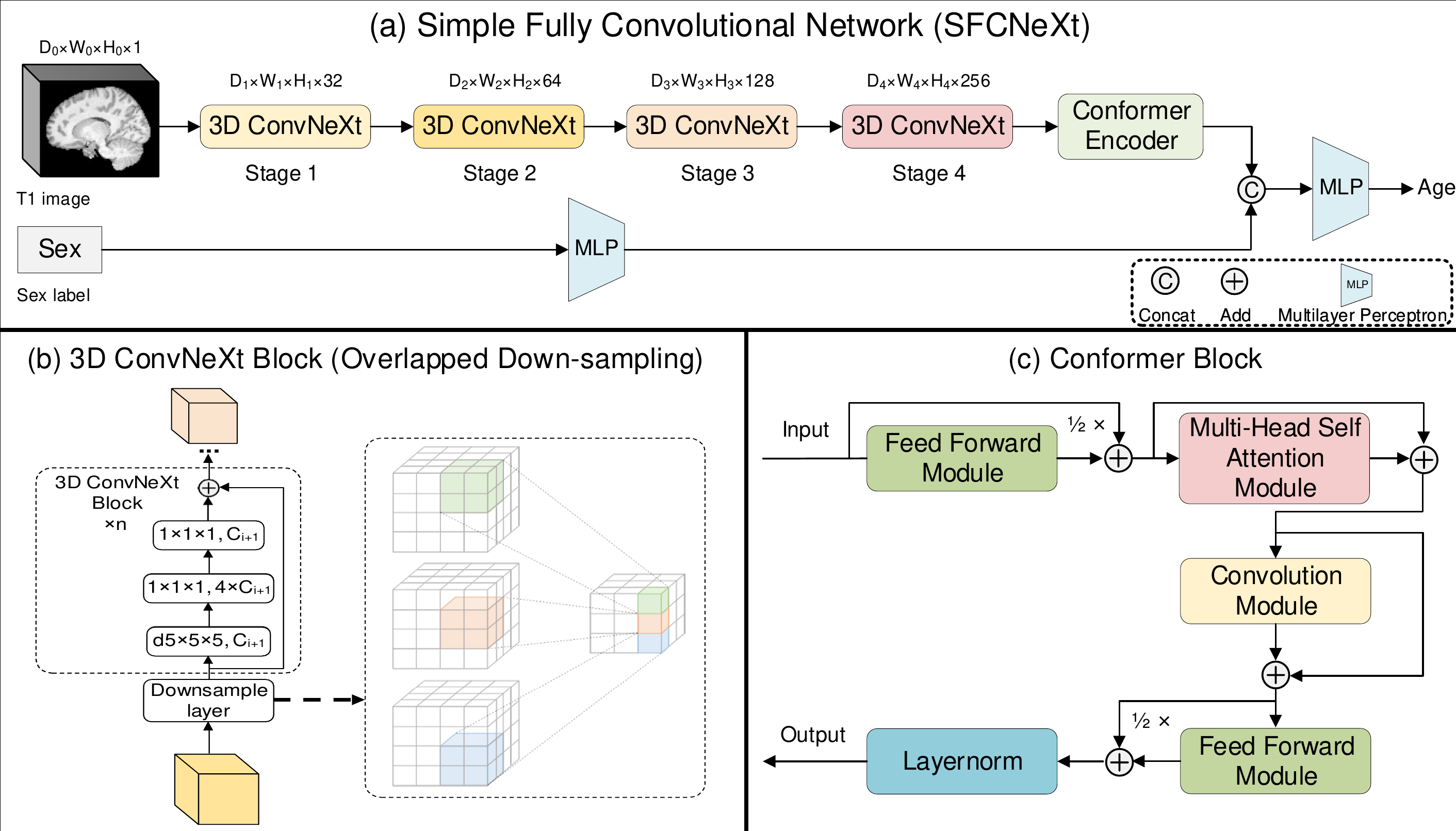}}
\caption{(a) The whole framework of the proposed SFCNeXt; (b) Overlapped 3D down-sampling; (c) Conformer block.} 
\label{framework}
\end{figure*}

The whole framework of the proposed SFCNeXt is displayed in Fig.~\ref{framework} (a). The SPEC consists of a backbone pathway and a branch pathway. The backbone pathway is a four-stage ConvNeXt network (i.e., each stage includes several ConvNeXt blocks) for down-sampling and aggregating 3D brain MRI features in a cascade manner. At the end of the backbone pathway, a conformer encoder~\cite{lu2019understanding,fu2022resource} module is employed for the adaptive encoding these high-level brain MRI features. The branch pathway is a simple multilayer perceptron (MLP) that mainly aims to feed sex features (i.e., male or female) into the model. Although controversial in previous studies~\cite{biskup2019sex}, we have found positive benefits of adding sex features in this study. After the concatenation of the two pathways, the brain age is obtained using the final MLP. The number of blocks in the four-stage ConvNeXt network is (1, 1, 3, 1), which significantly reduces the computing resources of the original ConvNeXt stage such as (3, 3, 9, 3)~\cite{liu2022convnet}. All of the ConvNeXt blocks in each stage of the SPEC use an overlapped working mode for 3D down-sampling (see Fig.~\ref{framework} (b)), which can significantly mitigate the feature damage problem caused by dividing each 3D brain into multiple non-overlapping 3D patches. Usually, a conformer encoder module (see Fig.~\ref{framework} (a)) contains $N$ conformer blocks, each of which contains two feed-forward (FF) modules sandwiching the multi-headed self-attention (MHSA) module and the convolution module (see Fig.~\ref{framework} (c)). Each conformer block consists of 4 residual connection modules and a layer normalization operator.

\vspace{-0.3cm}

\subsection{Hybrid Ranking Loss (HRL)}
Some previous studies have shown the benefits of ranking loss for improving the performance of brain age estimation~\cite{cheng2021brain,fu2022otfpf}. For example, \cite{cheng2021brain} adopts Spearman's rank correlation coefficient (SRCC) of chronological age to improve brain age estimation directly. For two subjects with ages $y_i$ and $y_j$, the age difference
loss can be seen as the MSE between the estimated brain age difference $\hat{y}_i-\hat{y}_j$ and the true age difference $y_i-y_j$. Suppose the $N_b$ denotes the number of paired samples $(i, j)$ in each batch, we can list the MSE loss $\mathcal{L}_{MSE}$ and age difference loss $\mathcal{L}_{\mathrm{d}}$ as follows, respectively:

\vspace{-0.3cm}

\begin{equation}\small\vspace{-2ex}
\mathcal{L}_{\mathrm{MSE}}=\frac{1}{N_b} \sum_{i=1}^{N_b}\left(\hat{y}_i-y_i\right)^2
\end{equation}

\begin{equation}\small
    \mathcal{L}_{\mathrm{d}}=\frac{1}{N_b} \sum_{i=1, j=1}^{N_b}\left(\left(\hat{y}_i-\hat{y}_j\right)-\left(y_i-y_j\right)\right)^2
\end{equation}
where the sign of the $\mathcal{L}_{\mathrm{d}}$ indicates the positive or negative difference of two ages. SRCC rank correlation can be defined as the PCC between the rank values of two variables:

\vspace{-0.1cm}

\begin{equation}\small
r_r=\frac{\operatorname{cov}(\operatorname{Rank}(\hat{y}), \operatorname{Rank}(y))}{\sigma_{\operatorname{Rank}(\hat{y})} \sigma_{\operatorname{Rank}(y)}}
\end{equation}
where $\operatorname{Rank}(y)$ is the rank operator, $\operatorname{cov}(\operatorname{Rank}(\hat{y}), \operatorname{Rank}(y))$ is the covariance of the two rank variables, and $\sigma_{\operatorname{Rank}(\hat{y})}$ and $\sigma_{\operatorname{Rank}(y)}$ are the standard deviations of the two rank variables. So the SRCC ranking loss can be simplified as:

\vspace{-0.3cm}

\begin{equation}\small\vspace{-1ex}
\mathcal{L}_{\mathrm{r}}=\sum_{i=1}^{N_b}\left(\operatorname{Rank}\left(\hat{y}_i\right)-\operatorname{Rank}\left(y_i\right)\right)^2
\end{equation}
As shown by~\cite{blondel2020fast}, computing SRCC is usually problematic: its derivatives are null or undefined, preventing gradient backpropagation.
A simple method based on comparing pairwise distances between variables can obtain the approximation of SRCCs, but may take $O\left(n^2\right)$ time~\cite{qin2010general}, which is not suitable for our intention in SFCNeXt.

To reduce the computational complexity, this paper feats the ranking as projections onto a permutahedron. Let $\boldsymbol{z}, \boldsymbol{w} \in \mathbb{R}^n$ and consider the linear program as:
\begin{equation}\small
\operatorname{argmax}_{\boldsymbol{\mu} \in \mathcal{P}(\boldsymbol{w})}\langle\boldsymbol{\mu}, \boldsymbol{z}\rangle
\end{equation}
then we can represent the $r(\boldsymbol{\theta})$ by setting $(\boldsymbol{z}, \boldsymbol{w})=(-\boldsymbol{\theta}, \boldsymbol{\rho})$. Here the $r(\boldsymbol{\theta})$ means rank values of $\theta$ and the $\mathcal{P}(\boldsymbol{w})$ denotes the permutahedron of $w$. $\theta$ and $\boldsymbol{\rho}$ denote permutations with linear program. Therefore, the $\Psi$-regularized soft ranking loss can be written as:

\vspace{-0.2cm}

\begin{equation}\small
r_{\varepsilon \Psi}(\boldsymbol{\theta}):=P_{\varepsilon \Psi}(-\boldsymbol{\theta}, \boldsymbol{\rho})=P_{\Psi}(-\boldsymbol{\theta} / \varepsilon, \boldsymbol{\rho})
\end{equation}
where $0<\varepsilon<\infty$, and $O(n \log n)$ time can be easily achieved, which facilitates our training. Finally, the total loss function $\mathcal{L}$ is:

\vspace{-1ex}
\begin{equation}\small
\mathcal{L}=\mathcal{L}_{\mathrm{MSE}}+\lambda_1 \mathcal{L}_{\mathrm{d}}+\lambda_2 \mathcal{L}_{r_{\varepsilon \Psi}(\boldsymbol{\theta})}
\end{equation}
where the $\lambda_1$ and $\lambda_2$ are regularization parameters.

\vspace{-0.3cm}

\section{Experiments and Results}\label{sec:results}
To ensure the stability of the results, the healthy cohort is randomly partitioned into three subsets: the training set (80\% of MRIs), the validation set (10\% of MRIs), and the test set (10\% of MRIs). The training-validation-test split is performed for total 10 times to conduct the 10-fold cross-validation procedure, which is commonly used in previous brain age estimation studies~\cite{bashyam2020mri,peng2021accurate}. All experiments are conducted using a single NVIDIA GeForce RTX 3090 GPU with PyTorch library, and the python version is 3.8.13. A group of recent 3D lightweight state-of-the-art (SOTA) models for brain age estimation, such as SFCN~\cite{peng2021accurate}, 3D ResNet~\cite{peng2021accurate,he2016deep} and their corresponding variants (e.g., with different layers, blocks, and optimizers), have been included in this study for comparison. For a fair comparison, all competing models are reproduced according to their released codes, and we further optimize them for better performance in our dataset. Sex labels are also used as covariates of these competing models for the improvement of performance.


\vspace{-0.4cm}

\subsection{Parameter evaluations of the SFCNeXt}
First, we evaluate the performance of the proposed SFCNeXt using different loss functions (MAE or MSE), batch sizes (from 4 to 20), initial learning rates (ILR) (from $1\times10^{-4}$ to  $1\times10^{-2}$), optimizers (Adam, AdamW or Adamax), attention heads (from 1 to 4) and conformer blocks (from 1 to 3). SFCNeXt (a) to SFCNeXt (p) in Table~\ref{Parameter_evaluation_table} depict SFCNeXt variants with different hyper-parameters. Firstly, in the top part of the Table~\ref{Parameter_evaluation_table}, we find that adopting MSE as loss and 8 as batch size can achieve the best MAE, PCC, and SRCC (SFCNeXt (e)). Then, in the middle part of the Table~\ref{Parameter_evaluation_table}, we note that combining the $1\times10^{-4}$ as ILR and Adamax (SFCNeXt (m)) can reduce the MAE to 3.683$\pm$0.009, increase the PCC to 0.859$\pm$0.001 and SRCC to 0.817$\pm$0.001, when with 2 attention heads and 3 conformer blocks in the conformer encoder module. In the bottom part of Table~\ref{Parameter_evaluation_table}, we vary the number of attention heads and Conformer blocks and demonstrate that 2 attention heads and 3 conformer blocks are the most suitable combinations for SFCNeXt. Here the SFCNeXt (m) and SFCNeXt (p) refer to the same model. Therefore, the final version of SFCNeXt adopts MSE as a loss function, 8 as batch size, $1\times{10}^{-3}$ as the initial learning rate, Adamax as an optimizer, 2 attention heads and 3 conformer blocks.

\vspace{-0.4cm}

\subsection{Comparison of SFCNeXt vs. other lightweight SOTA models}
\begin{figure}[ht]
 \centering
 \centerline{\includegraphics[width=8.8cm]{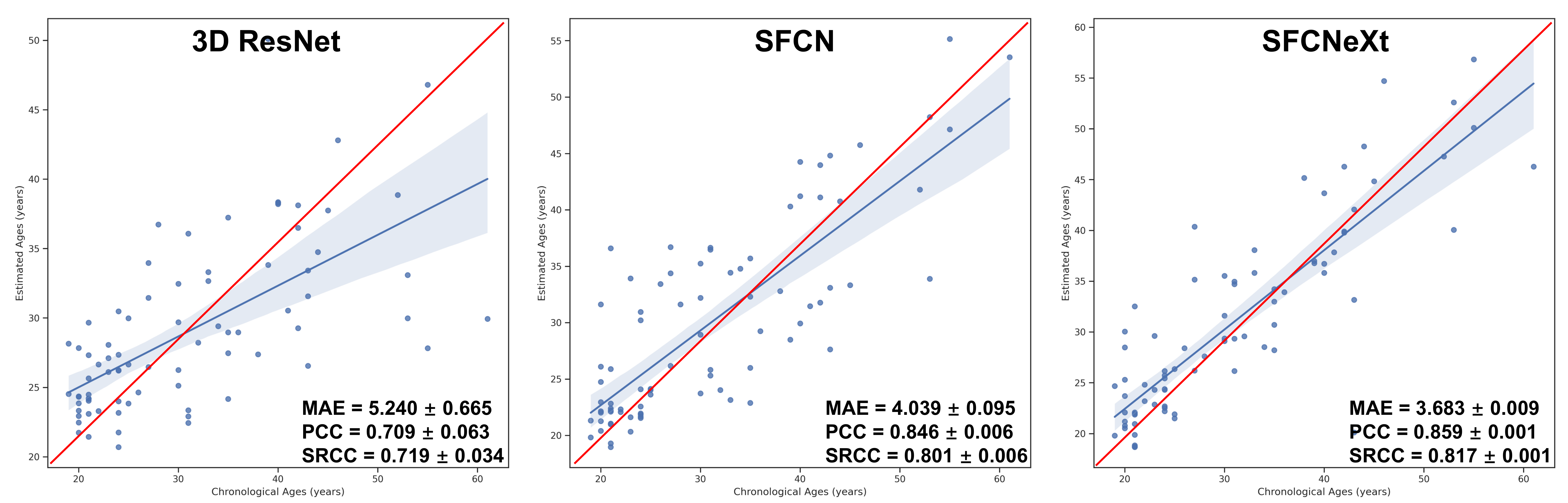}}
\caption{The scatter diagrams of estimated brain ages by 3D ResNet18, SFCN\_Adamax, and the SFCNeXt.The red lines in sub-graphs indicate the ideal estimation of y=x (i.e., the estimated brain age equals the chronological age), while the blue lines with CIs denote the actual estimation performance of each model. The smaller the angle between the two lines, the better the evaluation performance.}
\label{scatters}
\end{figure}

Table~\ref{Model_comparison_table} details the quantitative brain age estimation results for the test set by the SFCNeXt and its competing models (i.e. 3D ResNet, SFCN, and their corresponding variants). It is easy to note that the SFCNeXt outperforms all variants of 3D ResNet and SFCN. Fig.~\ref{scatters} depicts the scatter diagrams of the estimated brain ages versus the chronological brain ages based on the 3D ResNet18 (the best ResNet model), SFCN\_Adamax (the best SFCN model), and SFCNeXt. The SFCN\_Adam is found as the best model other than SFCNeXt. Compared with SFCN\_Adam, the SFCNeXt significantly reduces MAE by 9.7\%, PCC by 1.5\%, and SRCC by 2\%. Also, the SFCNeXt network consumes 6971MB of GPU memory, which is comparable to other lightweight competing models.


\begin{table}[ht]\scriptsize
 \centering
 \caption{Performance of brain age estimation using SFCNeXt with different hyper-parameters (bold denotes better results).}
 \label{Parameter_evaluation_table}
 \setlength{\tabcolsep}{0.55 mm}{
 \begin{threeparttable}
 \begin{tabular}{lcccccc}  
  \toprule   
  Model & Loss & Batch Size & \makecell[c]{MAE\\(Mean$\pm$STD)} & \makecell[c]{PCC\\(Mean$\pm$STD)} & \makecell[c]{SRCC\\(Mean$\pm$STD)} \\
  \midrule   
  SFCNeXt (a) & MAE & 4 & 4.396$\pm$0.322 & 0.802$\pm$0.005 & 0.790$\pm$0.012 \\  
  SFCNeXt (b) & MAE & 8 & 4.363$\pm$0.066 & 0.832$\pm$0.008 & 0.811$\pm$0.004 \\  
  SFCNeXt (c) & MAE & 20 & 8.369$\pm$0.063 & 0.293$\pm$0.046 & 0.275$\pm$0.051 \\ 
  SFCNeXt (d) & MSE & 4 & 4.514$\pm$0.116 & 0.814$\pm$0.011 & 0.772$\pm$0.007 \\  
  \textbf{SFCNeXt (e)} & \textbf{MSE} & \textbf{8} & \textbf{4.013$\pm$0.041} & \textbf{0.852$\pm$0.014} & \textbf{0.797$\pm$0.004} \\  
  SFCNeXt (f) & MSE & 20 & 5.489$\pm$0.131 & 0.774$\pm$0.017 & 0.781$\pm$0.003 \\ 
  \midrule    
  Model & ILR & Opmimizer & \makecell[c]{MAE\\(Mean$\pm$STD)} & \makecell[c]{PCC\\(Mean$\pm$STD)} & \makecell[c]{SRCC\\(Mean$\pm$STD)} \\
  \midrule   
  SFCNeXt (g) & $1\times10^{-3}$ & Adam & 4.013$\pm$0.041 & 0.852$\pm$0.014 & 0.787$\pm$0.004 \\  
  SFCNeXt (h) & $2\times10^{-3}$ & Adam & 4.467$\pm$0.061 & 0.810$\pm$0.009 & 0.772$\pm$0.008 \\  
  SFCNeXt (i) & $4\times10^{-3}$ & Adam & 6.497$\pm$1.986 & 0.671$\pm$0.184 & 0.658$\pm$0.145 \\  
  SFCNeXt (j) & $1\times10^{-3}$ & AdamW & 4.398$\pm$0.322 & 0.802$\pm$0.005 & 0.790$\pm$0.012 \\  
  SFCNeXt (k) & $2\times10^{-3}$ & AdamW & 4.426$\pm$0.143 & 0.802$\pm$0.017 & 0.771$\pm$0.026 \\  
  SFCNeXt (l) & $4\times10^{-3}$ & AdamW & 4.369$\pm$0.186 & 0.824$\pm$0.013 & 0.775$\pm$0.041 \\
  \textbf{SFCNeXt (m)} & \bm{$1\times10^{-3}$} & \textbf{Adamax} & \textbf{3.683$\pm$0.009} & \textbf{0.859$\pm$0.001} & \textbf{0.817$\pm$0.001} \\ 
  \midrule     

  Model & \makecell[c]{Attention\\Head} & \makecell[c]{Conformer\\Block} & \makecell[c]{MAE\\(Mean$\pm$STD)} & \makecell[c]{PCC\\(Mean$\pm$STD)} & \makecell[c]{SRCC\\(Mean$\pm$STD)} \\
  \midrule    
  SFCNeXt (n) & 2 & 1 & 3.919$\pm$0.075 & 0.854$\pm$0.006 & 0.814$\pm$0.002 \\  
  SFCNeXt (o) & 2 & 2 & 4.073$\pm$0.129 & 0.850$\pm$0.006 & 0.802$\pm$0.003 \\   
  \textbf{SFCNeXt (p)} & \textbf{2} & \textbf{3} & \textbf{3.683$\pm$0.009} & \textbf{0.859$\pm$0.001} & \textbf{0.817$\pm$0.001} \\
  SFCNeXt (q) & 4 & 3 & 3.951$\pm$0.017 & 0.852$\pm$0.006 & 0.808$\pm$0.004 \\ 
  \bottomrule  
 \end{tabular}
 \begin{tablenotes}[flushleft]
    \item MAE Mean Absolute Error; MSE Mean Square Error; ILR: Initial Learning Rate
 \end{tablenotes}
 \end{threeparttable}}
 \vspace{-15pt}
\end{table}


\begin{table}[t]\scriptsize
 \centering
 \caption{Brain age estimation results of SFCNeXt and other lightweight competing models.}
 \label{Model_comparison_table}
 \setlength{\tabcolsep}{1 mm}{
 \begin{threeparttable}
 \begin{tabular}{lccccc}  
\toprule 
Model &  \makecell[c]{MAE\\(Mean$\pm$STD)} & \makecell[c]{PCC\\(Mean$\pm$STD)} & \makecell[c]{SRCC\\(Mean$\pm$STD)} & \makecell[c]{GPU\\Consumption}\\
\midrule 
3D ResNet10 &  5.817$\pm$0.385 & 0.664$\pm$0.062 & 0.715$\pm$0.043 & 4125MB\\
3D ResNet18 & 5.240$\pm$0.665 & 0.709$\pm$0.063 & 0.719$\pm$0.034 & 4497MB\\
3D ResNet34 & 5.744$\pm$0.576 & 0.682$\pm$0.063 & 0.712$\pm$0.057 & 5133MB\\
3D ResNet152 & 5.514$\pm$0.334 & 0.714$\pm$0.034 & 0.743$\pm$0.009 & 8879MB\\
SFCN\_Adam & 4.624$\pm$0.204 & 0.800$\pm$0.026 & 0.787$\pm$0.026 & 6431MB\\
SFCN\_AdamW & 4.332$\pm$0.262 & 0.823$\pm$0.020 & 0.799$\pm$0.038 & 6431MB\\
SFCN\_Adamax &  4.039$\pm$0.095 & 0.846$\pm$0.006 & 0.801$\pm$0.006 & 6431MB\\
\textbf{SFCNeXt} & \textbf{3.683$\pm$0.009} & \textbf{0.859$\pm$0.001} & \textbf{0.817$\pm$0.001} & \textbf{6971MB}\\ 
\bottomrule
 \end{tabular}
 \end{threeparttable}}
\end{table}


\vspace{-0.4cm}

\begin{table}[ht]\scriptsize
 \centering
 \caption{Brain age estimation results of SFCNeXt and its corresponding variants.}
 \label{Ablation}
 \setlength{\tabcolsep}{1 mm}{
 \begin{threeparttable}
 \begin{tabular}{lcccc}  
\toprule 
Model & MAE (Mean$\pm$STD) & PCC (Mean$\pm$STD) & SRCC (Mean$\pm$STD) \\
\midrule 
SFCNeXt (1) & 4.177$\pm$0.067 & 0.834$\pm$0.004 & 0.801$\pm$0.031 \\
SFCNeXt (2) & 4.046$\pm$0.086 & 0.847$\pm$0.005 & 0.794$\pm$0.012 \\
SFCNeXt (3) & 4.216$\pm$0.082 & 0.848$\pm$0.001 & 0.811$\pm$0.002 \\
\textbf{SFCNeXt} & \textbf{3.683$\pm$0.009} & \textbf{0.859$\pm$0.001} & \textbf{0.817$\pm$0.001} \\ 
\bottomrule
 \end{tabular}
 \end{threeparttable}}
\end{table}


\vspace{-0.3cm}

\subsection{Ablation study of the SFCNeXt}
To demonstrate the rationality of the entire SFCNeXt framework, we also perform ablation experiments for SFCNeXt versus its variants. For example, the "SFCNeXt (1)", "SFCNeXt (2)," and "SFCNeXt (3)" in Table~\ref{Ablation} represent three variants of the SFCNeXt: "SFCNeXt w/o Sex Label," "SFCNeXt w/o Conformer Encoder" and "SFCNeXt Using Original ConvNeXt Stage (3, 3, 9, 3)". Results in Table~\ref{Ablation} can help us understand the superiority of SFCNeXt relative to all of its variants quantitatively. We can note that the use of a conformer block (SFCNeXt (2)) can reduce the MAE by 0.363, increasing the PCC by 1.4\% and SRCC by 2.9\%. The sex label is also very important for the performance gain: without the sex label (see the SFCNeXt (1)), the MAE degenerates to 4.177, PCC decreases to 0.834, and SRCC decreases to 0.801. All ablation experiments have shown the rationality of adopting these modules in SFCNeXt itself.


\vspace{-0.3cm}

\section{Conclusion}
\label{sec:discuss}
This study proposes the SFCNeXt as a lightweight end-to-end neural network architecture for effective brain age estimation with T1 MRIs. We conduct widespread experiments on a multi-site small-sized healthy cohort with biased age distribution and achieve the MAE of 3.683, PCC of 0.859, and SRCC of 0.817. Firstly, we evaluate the different combinations of SFCNeXt's parameters, and justify the parameter settings of SFCNeXt. Secondly, we compare the SFCNeXt with current lightweight SOTA models, and the quantitative comparison results of SFCNeXt show its superiority in classical metrics, such as MAE, PCC, and SRCC. Thirdly, the ablation study is performed, and all results demonstrate the rationality and necessity of all modules adopted in the SFCNeXt. To our knowledge, this is the first lightweight model specially designed for small sample brain age estimation. In conclusion, the SFCNeXt consumes comparable GPU resources to other lightweight models, is easy to deploy, and is suitable for medical situations which own insufficient subjects. In the future, it can be implemented for MRIs of brain disease or disorder groups to predict the longitudinal development trajectory and help to create an early intervention strategy. 

\section{Compliance with Ethical Standards}
\label{sec:ethical}

This research study was conducted retrospectively using human subject data available in open access by~\cite{yahata2016small,tanaka2021multi}. Ethical approval was not required, as confirmed by the license attached with the open-access data.

\section{Acknowledgements}
\label{sec:acknowledgments}

This work was supported in part by the Zhejiang Provincial Innovation Team Project (No. 2020R01001) and in part by the Innovation Fund of the Department of Education of Gansu Province (2022B-023).




\bibliographystyle{IEEEbib}
\fontsize{9pt}{10pt}
\selectfont
\bibliography{strings,refs}

\end{document}